# Design Principles & Issues for Gaze and Pinch Interaction

Ken Pfeuffer, Assistant Professor, Aarhus University                18. January 2024

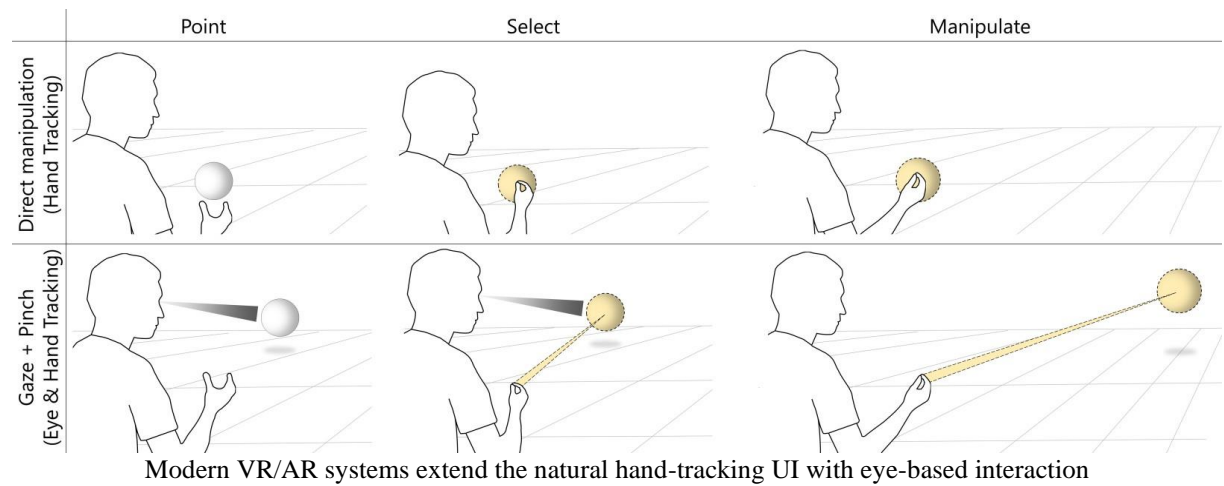

Modern VR/AR systems extend the natural hand-tracking UI with eye-based interaction

Controllers, hand gestures, eye movements, and voice: many ways to click buttons in virtual reality environments. What about: glance at a UI object with your eyes, then simply pinch with your fingers to activate it. Apple innovates with the first wide adoption of this interaction style for their Vision Pro spatial computer. As well the Hololens 2 and Magic Leap offered similar functionalities. But Apple, renowned for stellar product design, may nail it. Early users are raving about the [mind-blowing](#) and [telepathic](#) technology.

To shed light on the interaction design, we present 5 design principles and 5 design issues. These are based on human-computer interaction research, mostly the paper "Gaze + Pinch Interaction in Virtual Reality" presented at the 2017 Spatial User Interfaces symposium. We'll see how much Apple has considered the scientific roots when we get our hands on it!

# Design Principles

## 1. Division of labor: The eyes select, the hands manipulate

Our eyes' natural role involves indicating points of interest, and we can easily look at any point at will. In contrast, the hands are adept at physical manipulation through the interplay of finger movement and hand posture. Use a clear separation of concerns: the eyes perform selection tasks, the hands do the actual work. This avoids the pitfalls of (i) overloading the eyes with motor control tasks —here you only actively "use" the eyes to select, and (ii) physical fatigue —gaze pointing minimizing the hands' physical motion needs.

The hands, then, make indirect gestures. This is similar to a controller in having the ability to interact at a distance, but now with intuitive pinch gestures. Actually, there are hands-only techniques for selection and manipulation, such as the defaults of the Meta Quest series and Hololens 2. Yet, assigning both selection and manipulation to the hand can be susceptible to hand jitter issues (the Heisenberg problem), and studies (a, b) showed that combining eyes and hands leads to improved performance and comfort.

Example: https://twitter.com/KenPfeuffer/status/1674198938650390528

## 2. Minimalistic multimodal timing

A single moment in time matters for the eyes—the moment that the index finger and thumb have first contact, one has to fixate on the desired target. Afterward, whether it's a drag, pan, or zoom gesture, the eyes relinquish control. This affords the freedom to inspect the surroundings independently and avoids accidental actions by eye input ('Midas Touch' problem). For instance, in drag & drop after selection, one can freely look around to locate the destination for the dragged object and follow with the hand via indirect control.

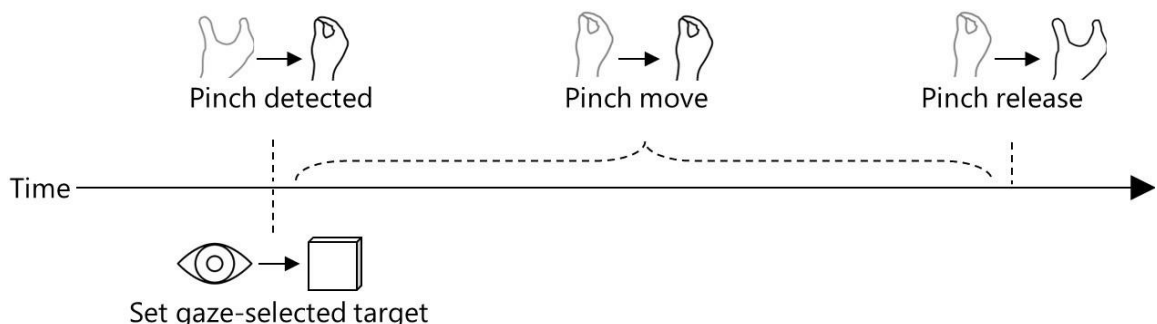
Eye-tracking as input is only active the moment that a pinch gesture is registered to avoid erratic behaviours.

The hands alone allow you to point without continuously monitoring the target. Gaze + Pinch inverses the relationship as the eyes must be on the target but the hands can be anywhere. Employing gaze minimally facilitates a quick adaptation of this new relationship. Beginners may find themselves more attentive to ensure their gaze is on the UI element until receiving the right feedback. More experienced users may swiftly execute a Gaze + Pinch command even without having fully perceived the target and its selection feedback.

## 3. Flexible gesture support

Hand gestures are in control of virtual objects acquired by the eyes. This can flexibly extend to all atomic classes of the hand-based manipulation—one vs. two-handed interactions, and single vs. multiple target manipulation. Users can seamlessly shift between 1 or two objects and hands by simply re-engaging pinch gestures as desired. This is inherited from the default hand-tracking gestures— but the indirection by gaze renders all those basic hand actions extremely lightweight and flexible.

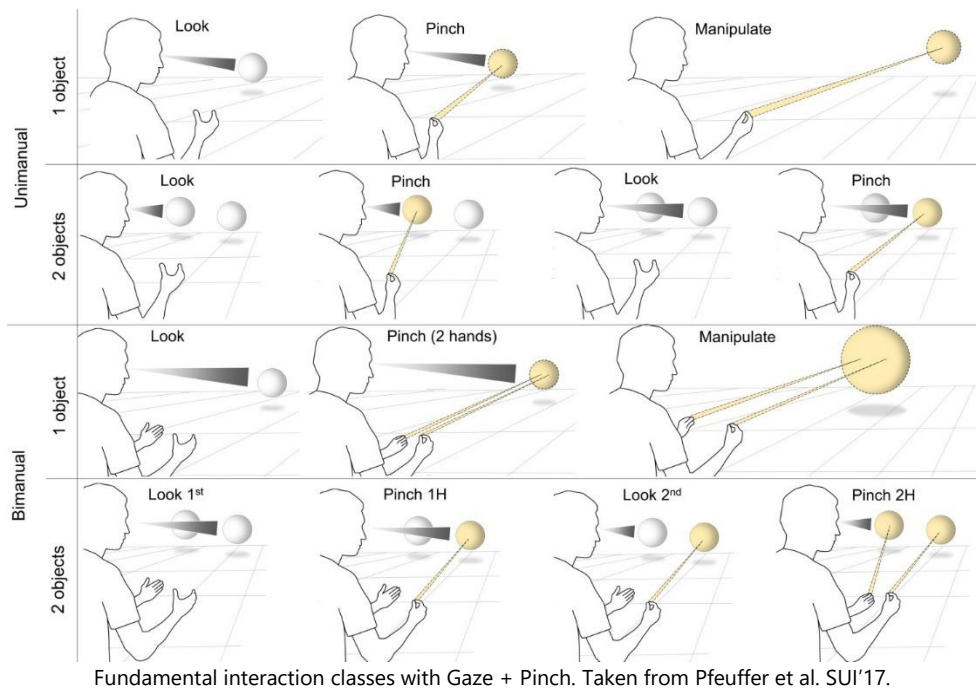

Fundamental interaction classes with Gaze + Pinch. Taken from Pfeuffer et al. SUI'17.

## 4. Infallible eyes

Our hands adhere to a speed-accuracy trade-off: faster means less accurate. In contrast, the eyes can fixate on a target in almost instant time, with constant accuracy given by the eye-tracking sensor. From a user's perspective, the eyes are infallible: hand pointing can miss or overshoot a target, but the eyes can't miss as we are either on-target or we look elsewhere. The content design may need to be adjusted, to account for the altered attention of the user during the computer interaction tasks, as well as how UIs support undo and error-recovery measures.

Large buttons invite wandering around with the eyes, potentially leading to outliers. Drawing the most salient parts to the center of the button will be welcomed by the selection mechanism, and a generous buffer space around targets makes outliers less impactful. If smaller targets are required, hand-refinement and cursor extensions can be integrated with Gaze + Pinch, although departing from the original simplicity.

Example: https://twitter.com/KenPfeuffer/status/1747662769984287005

### 5. Compatibility with hand-based interfaces

The motto is to get the best of both worlds. Gaze + Pinch can also work together with the direct gestures in nearspace. This is possible via [mode-switching methods](#) that use time and space multiplexing of the inputs. For time-multiplexing, imagine a user opens an app with a menu through Gaze + Pinch, which activates the app UI in front of the user. The user switches to direct touch gestures to scroll the app's content. In space multiplexing, picture holding a menu with one hand while using Gaze + Pinch commands with the other, enabling direct and indirect inputs at the same time for on-hand virtual menus. Hand menus usually position menus off the hand to prevent hand-tracking interference. Indirect gestures are spatially separated, avoiding hand overlap. It's one of the intriguing outcomes when UI systems support transitions between complementary modes of interaction.

Example: https://twitter.com/KenPfeuffer/status/1706754929438654798

## Some issues

### 1. (Un)Learning

When we want to get something, we intuitively move toward it. Gaze + Pinch commands don't need this movement anymore, which can be considered almost counter-intuitive. Hence it is important to balance the advantage of using a new way that requires re-thinking of the action process, over just using the hands without the eyes. In a sense, Gaze + Pinch does not mean learning a new way of interaction — it's about unlearning the common way: don't move your hands, just confirm right where you are with your hand.

### 2. Early- and late-triggers

A [known problem](#) of multimodal interaction is that the eyes may leave the target before a manual command is registered, or the command is issued just before the gaze lands on the target. It's possible to have a predictive and generous timing, e.g., using the last fixation (~200–300ms of a stable gaze), rather than the current gaze coordinate (see [fixation detection methods](#)). If the early or late trigger frequency is known for a specific UI, the timing can be adapted.

### 3. Control-display ratio

With direct manipulation the dragged object translates 1:1 with the hand. With Gaze + Pinch, moving a distant object at a 1:1 mapping makes it feel very slow. What one can do is amplify the speed in the transfer function with the object distance.
Or, use visual angle to determine dragging speed as a distance-independent metric:

If your hand moves by 5 degrees in your FoV, the remote object corresponds with 5 degree motion. This works well for objects at a distance but may be confusing when targets are near — here the UI can revert to a 1:1 transfer function.

Example: https://twitter.com/KenPfeuffer/status/1675851939286855685

### 4. Drag & drop sequences

In regular interactions, dropping something means your hand is right there for you to pick it up again. But with Gaze + Pinch, you can look away after dropping and finding it again means you have to look back. This can be a hassle, especially if turning your body is involved. So, when designing interfaces, it's crucial to think about what type of tasks are supported — ideally, most tasks require only a single action to finish drag & drop, and for sequences consider potential enhancements. Hand tracking systems can make sure not to disengage the target from the control of a pinch gesture if just briefly undetected, to keep dragging robust and avoid object loss.

### 5. Continuous eye-selection

With a two-handed pinch-to-zoom gesture, one can use the eyes continuously to refine the zooming pivot. Doing this will put more responsibility on the eyes, which for zooming and related tasks can feel natural. Less panning actions can be the result, if the end of a zooming gesture means that we zoomed closer to where we wanted to than when we used the hands.

## Summary

The interaction design for the eyes and hands is a novel space that is gradually gaining momentum. Gaze + Pinch or the Vision Pro UI are examples with a mix of underlying design principles and issues, that may just hit the right balance for transforming the ways we have been using our hands after all. It will be exciting to see how the principles will apply in practice, and what else may be ahead in the coming years for XR UX. To read more, check out novel eye-hand concepts such as Gaze-Hand Alignment, Gaze + Pinch to mobile hand menus, a short video for Gaze + Pinch applications, and more below.

# Further material

The article is primarily based on the author's experiences, notes, and thoughts, with some knowledge of the human-computer interaction (HCI) research field.

## Author & colleagues' research

- Paper on Gaze + pinch interaction in virtual reality (SUI '17). By Ken Pfeuffer, Benedikt Mayer, Diako Mardanbegi, and Hans Gellersen. [doi](doi), [pdf](pdf), [video](video), [talk](talk)
- Eye-Hand Symbiosis project: https://kenpfeuffer.com/eye-hand-symbiosis-what-guide/
- The XI team at Aarhus University: https://kenpfeuffer.com/extended-interactionxi-team/
- The GEMINI project studies eye and body motion: https://gemini-erc.eu/

## Related research literature

Important precursory research has been done, among many others, by Sophie Stellmach et al. (CHI'12–13) and Jayson Turner et al. (INTERACT'13, MUM'13, ETRA'14, CHI'15), on gaze and touch gesture for large displays. Zhai et al's MAGIC (CHI'99), Jacob's What You Look At Is What You Get (CHI'90), and Bolt's Gaze-orchestrated dynamic windows (SIGGRAPH'81) are early influential entries in the field.

## Design guidelines and tutorials

- For Microsoft's Mixed Reality Toolkit (MRTK): https://learn.microsoft.com/en-us/windows/mixed-reality/mrtk-unity/mrtk3-overview/architecture/interactors
- For the Meta Quest Pro (by Dilmer Valecillos), for Unity https://youtu.be/2xWhBKn7Wp0?si=Aq2FVPdkPYVv-tm8, for ShapesXR https://www.youtube.com/watch?v=YQOHliIeuMY&ab_channel=ShapesXR
- Apple Vision Pro "Designing for visionOS": https://developer.apple.com/design/human-interface-guidelines/designing-for-visionos

## Related articles

- Eyes & hands in AR: A sci-fi-inspired mobile UI research: https://www.linkedin.com/pulse/eyes-hands-ar-sci-fi-inspired-mobile-ui-research-ken-pfeuffer/?utm_source=rss&utm_campaign=articles_sitemaps
- Summary of Eye-Hand HCI Research in the Past 10 Years for Apple's Vision OS: https://medium.com/@cybrain/summary-of-eye-hand-hci-research-in-the-past-10-years-for-apples-vision-os-6585ba52716c
- Fitts's Law meets Apple's Vision Pro—Ergonomic design challenges from mobile to spatial computing: https://medium.com/@jim-ekanem/fittss-law-meets-apple-s-vision-pro-ergonomic-design-challenges-from-mobile-to-spatial-6c8994ecf975
- Gaze-Based Interaction: 30 Years in Retrospect and Future Outlook https://medium.com/@cybrain/gaze-based-interaction-30-years-in-retrospect-and-future-outlook-af923a32a921